\documentclass[english,aps,prl,twocolumn,showpacs]{revtex4}
\usepackage[T1]{fontenc}
\usepackage[latin9]{inputenc}
\usepackage{amsmath}
\usepackage{amssymb}
\usepackage{graphicx}
\usepackage{esint}
\usepackage{epstopdf}

\makeatletter
\@ifundefined{textcolor}{}
{%
 \definecolor{BLACK}{gray}{0}
 \definecolor{WHITE}{gray}{1}
 \definecolor{RED}{rgb}{1,0,0}
 \definecolor{GREEN}{rgb}{0,1,0}
 \definecolor{BLUE}{rgb}{0,0,1}
 \definecolor{CYAN}{cmyk}{1,0,0,0}
 \definecolor{MAGENTA}{cmyk}{0,1,0,0}
 \definecolor{YELLOW}{cmyk}{0,0,1,0}
}

\usepackage{upgreek}

\makeatother

\usepackage{babel}
\begin{document}

\title{High Fidelity Quantum Gates for Trapped Ions under Micromotion}

\author{C. Shen and L.-M. Duan}

\affiliation{Department of physics, University of Michigan, Ann Arbor, MI 48109}

\affiliation{Center for Quantum Information, IIIS, Tsinghua University, Beijing
100084, China}
\begin{abstract}
Two or three dimensional Paul traps can confine a large number of
ions forming a Wigner crystal, which would provide an ideal architecture
for scalable quantum computation except for the micromotion, an issue
that is widely believed to be the killer for high fidelity quantum
gates. Surprisingly, here we show that the micromotion is not an obstacle
at all for design of high fidelity quantum gates, even though the
magnitude of the micromotion is significantly beyond the requirement
of the Lamb-Dicke condition. Through exact solution of the quantum
Mathieu equations, we demonstrate the principle of the gate design
under micromotion using two ions in a quadrupole Paul trap as an example.
The proposed micromotion quantum gates can be extended to the many
ion case, paving a new way for scalable trapped ion quantum computation. 
\end{abstract}


\maketitle
Trapped ions constitute one of the most promising systems for realization
of quantum computation \cite{1}. All the quantum information processing
experiments so far are done in linear Paul traps, where the ions form
a one-dimensional (1D) crystal along the trap axis \cite{1,2,3,4}.
In this configuration, the external radio-frequency (r.f.) Paul trap
can be well approximated by a static trapping potential with negligible
micromotion, which is believed to be critical for design of high fidelity
quantum gates. However, in term of scalability, the linear configuration
is not the optimal one for realization of large scale quantum computation:
first, the number of ions in a linear trap is limited \cite{5}; and
second, the linear configuration is not convenient for realization
of fault-tolerant quantum computation. The effective qubit coupling
in a large ion chain is dominated by the dipole interaction, which
is only good for short-range quantum gates because of its fast decay
with distance. In a linear chain with short range quantum gates, the
error threshold for fault tolerance is very tough and hard to be met
experimentally \cite{6,raussendorf_fault-tolerant_2007}.

From a scalability point of view, two (2D) or three dimensional (3D)
Paul traps would be much better for quantum computation compared with
a linear chain. In a 2D or 3D trap, one can hold a large number of
qubits with a high error threshold for fault tolerance, in the range
of a percent level, even with just the nearest neighbor quantum gates
\cite{raussendorf_fault-tolerant_2007}. Thousands to millions of
ions have been successfully trapped to form 2D or 3D Wigner crystals
in a Paul trap \cite{2D_crystal}. However, there is a critical problem
to use this system for quantum computation, i.e., the micromotion
issue. In the 2D or 3D configuration, micromotion cannot be compensated,
and the magnitude of micromotion for each ion can be significantly
beyond the optical wavelength (i.e., outside of the Lamb-Dicke region).
As the micromotion is from the driving force of the Paul trap, it
cannot be laser cooled. The messy and large-magnitude micromotion
well beyond the Lamb-Dicke condition is believed to be a critical
hurdle for design of entangling quantum gate operations in this architecture.

In this paper, we show that the micromotion surprisingly is not an
obstacle at all for design of high-fidelity quantum gates. When the
ions form a crystal in a time-dependent Paul trap, they will be described
by a set of Mathieu equations. We solve exactly the quantum Mathieu
equations in general with an inhomogeneous driving term and find that
the micromotion is dominated by a well-defined classical trajectory
with no quantum fluctuation. This large classical motion is far outside
of the Lamb-Dicke region, however, it does not lead to infidelity
of quantum gates if it is appropriately taken into account in the
gate design. The quantum part of the Mathieu equation is described
by the secular mode with a micromotion correction to its mode function.
This part of motion still satisfies the Lamb-Dicke condition at the
Doppler temperature, which is routine to achieve for experiments.
We use two ions in a quadrupole trap, which have large micromotion,
as an example to show the principle of the gate design, and give the
explicit gate scheme both in the slow and the fast gate regions using
multi-segment laser pulses \cite{zhu_arbitrary-speed_2006,zhu_trapped_2006},
with the intrinsic gate infidelity arbitrarily approaching zero under
large micromotion. We finally give a breif discussion of the general
procedure of the gate design under micromotion, which in principle
can work for any number of ions, with important implication for large-scale
quantum computation.

To illustrate the general feature of micromotion in a Paul trap and
the principle of the gate design under micromotion, we consider a
three-dimensional (3D) anisotropic quadrupole trap with a time dependent
potential $\Phi(x,y,z)=\left(U_{0}+V_{0}\cos\left(\Omega_{T}t\right)\right)\left(\frac{x^{2}+y^{2}-2z^{2}}{d_{0}^{2}}\right)\equiv\alpha(t)(x^{2}+y^{2}-2z^{2})$
from an electric field oscillating at the r.f. $\Omega_{T}$, where
$U_{0},V_{0}$ are voltages for the d.c. and a.c. components and $d_{0}$
characterizes the size of the trap. We choose a positive $U_{0}$
to reduce the effective trap strength along the $z$ direction so
that the two ions align along the $z$-axis. Since the motions in
different directions do not couple to each other under quadratic expansion,
we focus our attention on the $z$ direction. The total potential
energy of two ions (each with charge $e$ and mass $m$) is 
\begin{equation}
V(z_{1},z_{2})=-2e\alpha(t)\left(z_{1}^{2}+z_{2}^{2}\right)+\frac{e^{2}}{4\pi\epsilon_{0}\left\vert z_{1}-z_{2}\right\vert }.\label{1}
\end{equation}
Define center-of-mass (CM) coordinate $u_{\text{cm}}=(z_{1}+z_{2})/2$
and relative coordinate $u_{\text{r}}=z_{1}-z_{2}$. Without loss
of generality, we assume $u_{\text{r}}>0$ and its average $\bar{u}_{\text{r}}=u_{0}$.
We assume the magnitude of the ion motion is significantly less than
the ion separation, which is always true for the ions in a crystal
phase. The Coulomb interaction can then be expanded around the average
distance $\bar{u}_{\text{r}}$ up to the second order of $\left\vert u_{\text{r}}-u_{0}\right\vert $.
Under this expansion, the total Hamiltonian $H=p_{\text{cm}}^{2}/4m+p_{\text{r}}^{2}/m+V(z_{1},\, z_{2})$
is quadratic (although time-dependent) in terms of the coordinate
operators $u_{\text{cm}},u_{\text{r}}$ and the corresponding momentum
operators $p_{\text{cm}}=p_{1}+p_{2}$, $p_{\text{r}}=\left(p_{1}-p_{2}\right)/2$.
The Heisenberg equations under this Hamiltonian $H$ yield the following
quantum Mathieu equations respectively for the coordinate operators
$u_{\text{cm}}$ and $u_{\text{r}}$ 
\begin{equation}
\frac{d^{2}u_{\text{cm}}}{d\xi^{2}}+\left(a_{\text{cm}}-2q_{\text{cm}}\cos\left(2\xi\right)\right)u_{\text{cm}}=0\label{eq:EOM-CM}
\end{equation}
\begin{equation}
\frac{d^{2}u_{\text{r}}}{d\xi^{2}}+\left(a_{\text{r}}-2q_{\text{r}}\cos\left(2\xi\right)\right)u_{\text{r}}=f_{0}\label{eq:EOM-REL}
\end{equation}
where the dimensionless parameters $a_{\text{cm}}=-16eU_{0}/\left(md_{0}^{2}\Omega_{T}^{2}\right)$,
$a_{\text{r}}=a_{\text{cm}}+4e^{2}/\left(\pi\epsilon_{0}mu_{0}^{3}\Omega_{T}^{2}\right)$,
$q_{\text{\text{cm}}}=q_{\text{r}}=8eV_{0}/\left(md_{0}^{2}\Omega_{T}^{2}\right)$
and the dimensionless time $\xi=\Omega_{T}t/2$. The driving term
$f_{0}=6e^{2}/\left(\pi\epsilon_{0}mu_{0}^{2}\Omega_{T}^{2}\right)$.
The quantum operators $u_{\text{cm}}$ and $u_{\text{r}}$ satisfy
the same form of the Mathieu equations (except for the driving term
$f_{0}$) as for the classical variables. As these equations are linear,
we can use the solutions known for the classical Mathieu equation
to construct a quantum solution that takes into account of the quantum
fluctuation.

It is well known that the solution to the classical Mathieu equation$\frac{d^{2}}{d\xi^{2}}v+\left(a-2q\cos\left(2\xi\right)\right)v=0$
is a combination of Mathieu sine $S(a,\, q,\,\xi)$ and Mathieu cosine
$C(a,\, q,\,\xi)$ functions, which reduce to the conventional sine
and cosine functions when micromotion is neglected \cite{mclachlan_mathieu_1947}.
The solution to a homogeneous quantum Mathieu equation $\frac{d^{2}}{d\xi^{2}}\hat{u}+\left(a-2q\cos\left(2\xi\right)\right)\hat{u}=0$
can be described using the reference oscillator technique \cite{ref_oscillator}.
From the classical solution $v$ and the quantum operator $\hat{u}$,
one can introduce the following annihilation operator of a reference
oscillator (remember that $\xi=\Omega_{T}t/2$ is the dimensionless
time) 
\begin{equation}
\hat{a}(t)=\sqrt{\frac{m}{2\hbar\omega}}i\left(v(t)\dot{\hat{u}}(t)-\dot{v}(t)\hat{u}(t)\right),\label{eq:annihlation}
\end{equation}
where $\omega$ is a normalization constant typically taken as the
secular motion frequency of the corresponding Mathieu equation. In
addition, we impose the initial condition for $v(t)$ with $\left.v(t)\right\vert _{t=0}=1$
and $\left.\dot{v}(t)\right\vert _{t=0}=i\omega$. The position operator
$\hat{u}(t)$ and its conjugate momentum $\hat{p}(t)\equiv m\dot{\hat{u}}(t)$
satisfy the commutator $\left[\hat{u}(t),\hat{p}(t)\right]=i\hbar$.
From the above definition, one can easily check that $\frac{d}{dt}\hat{a}(t)\propto v\frac{d^{2}}{d\xi^{2}}\hat{u}-\hat{u}\frac{d^{2}}{d\xi^{2}}v=0$,
so $\hat{a}(t)\equiv\hat{a}$ is a constant of motion. Furthermore,
$\hat{a}$ and $\hat{a}^{\dagger}$ satisfy the standard commutator
\[
\left[\hat{a},\hat{a}^{\dagger}\right]=\left(m/2\hbar\omega\right)\left(i\hbar/m\right)\left.\left(v(t)\dot{v}^{\ast}(t)-v^{\ast}(t)\dot{v}(t)\right)\right\vert _{t=0}=1.
\]
When micromotion is neglected, $v(t)=e^{i\omega t}$ and $\hat{a}$
reduces to the annihilation operator of a harmonic oscillator. in
the presence of micromotion, $v(t)=C(a,\, q,\,\xi)+iS(a,\, q,\,\xi)$.
The solution to the position operator $\hat{u}$ takes the form 
\begin{equation}
\hat{u}(t)=u_{0}\left(v^{\ast}(t)\hat{a}+v(t)\hat{a}^{\dagger}\right)\label{5}
\end{equation}
where $u_{0}\equiv\sqrt{\hbar/2m\omega}$ is the oscillator length.

The above solution gives a complete description of the center-of-mass
motion with the operator 
\begin{equation}
u_{\text{cm}}(t)=u_{0\text{cm}}\left(v_{\text{cm}}^{\ast}(t)\hat{a}_{\text{cm}}+v_{\text{cm}}(t)\hat{a}_{\text{cm}}^{\dagger}\right),\label{6}
\end{equation}
where $u_{0\text{cm}}\equiv\sqrt{\hbar/4m\omega_{\text{cm}}}$ and
$\omega_{cm}$ is the secular frequency of the center of mass mode.
The relative motion $u_{\text{r}}$ satisfies the inhomogeneous quantum
Mathieu equation (3). To solve it, we let $u_{\text{r}}=u_{\text{r}}^{\prime}+\bar{u}_{\text{r}}$,
where $u_{\text{r}}^{\prime}$ is an operator that inherits the commutators
for $u_{\text{r}}$ and satisfies the homogenous quantum Mathieu equation
and $\bar{u}_{\text{r}}$ is a classical variable corresponding to
a special solution of the Mathieu equation $\frac{d^{2}\bar{u}_{\text{r}}}{d\xi^{2}}+\left(a_{\text{r}}-2q_{\text{r}}\cos\left(2\xi\right)\right)\bar{u}_{\text{r}}=f_{0}$.
The special solution $\bar{u}_{\text{r}}$ can be found through the
series expansion $\bar{u}_{\text{r}}=f_{0}\sum_{n=0}^{+\infty}c_{n}\cos(2n\xi)$,
where the expansion coefficients $c_{n}$ satisfy the recursion relations
$a_{\text{r}}c_{0}-q_{r}c_{1}=1$ and $c_{n}=D_{n}\left(c_{n-1}+c_{n+1}+c_{0}\delta_{n,1}\right)$
for $n\geq1$ with $D_{n}\equiv-q_{r}/\left(4n^{2}-a_{r}\right)$.
When $a_{r}\ll1$ and $q_{\text{r}}\ll1$, which is typically true
under real experimental configurations, $c_{n}$ rapidly decays to
zero with $\left\vert c_{n+1}/c_{n}\right\vert \approx q_{\text{r}}/4\left(n+1\right)^{2}$
and we can keep only the first few terms in the expansion and obtain
an approximate analytical expression for $\bar{u}_{\text{r}}$ \cite{supplementary}.
The complete solution of $u_{\text{r}}$ is then given by 
\begin{equation}
u_{\text{r}}(t)=u_{0\text{r}}\left(v_{\text{r}}^{\ast}(t)\hat{a}_{\text{r}}+v_{\text{r}}(t)\hat{a}_{\text{r}}^{\dagger}\right)+\bar{u}_{\text{r}}(t),\label{7}
\end{equation}
where $u_{0\text{r}}\equiv\sqrt{\hbar/m\omega_{\text{r}}}$ and $\omega_{r}$
is the secular frequency of the relative mode.

Now we show how to design high fidelity quantum gates under micromotion.
To perform the controlled phase flip (CPF) gate, we apply laser induced
spin dependent force on the ions, with the interaction Hamiltonian
described by \cite{zhu_trapped_2006} 
\begin{equation}
H=\sum_{j=1}^{2}\hbar\Omega_{j}\cos\left(k_{\delta}z_{j}+\mu_{\delta}t+\phi_{j}\right)\sigma_{j}^{z}.\label{8}
\end{equation}
where $k_{\delta}$ is the wave vector difference of the two Raman
beams along the $z$ direction, $\mu_{\delta}$ is the two-photon
Raman detuning, $\Omega_{j}$ (real) is the Raman Rabi frequency for
the ion $j$, and $\phi_{j}$ is the corresponding initial phase.
In terms of the normal modes, the position operators $z_{j}=u_{\text{cm}}-(-1)^{j}u_{\text{r}}/2$,
where $u_{\text{cm}},\, u_{\text{r}}$ are given by Eqs. (6) and (7).
We introduce three Lamb-Dicke parameters, $\eta_{\text{cm}}\equiv k_{\delta}u_{0\text{cm}}$
for the CM mode, $\eta_{\text{\text{r}}}\equiv k_{\delta}u_{0\text{r}}/2$
for the relative mode, and $\eta_{\text{mm}}\equiv k_{\delta}\bar{u}_{\text{r}}/2$
for pure micromotion. Under typical experimental configurations, $\eta_{\text{cm}}\sim\eta_{\text{r}}\ll1$.
The parameter $\eta_{\text{mm}}$ is a classical variable that oscillates
rapidly with time by multiples of the micromotion frequency $\Omega_{T}$.
In Fig. 1(a), we show a typical trajectory of $\eta_{\text{mm}}\left(t\right)$.
The magnitude of variation of $\eta_{\text{mm}}$ is considerably
larger than $1$. In Fig. 1(b), we also plot the function $v_{\text{cm}}(t)$,
which is dominated by the oscillation at the secular motion frequency
$\omega_{\text{cm}}$ with small correction from the micromotion.
The magnitude of $v_{\text{cm}}(t)$ is bounded by a constant slightly
larger than 1. The function $v_{r}(t)$ has very similar behavior
except that $\omega_{\text{cm}}$ is replaced by $\omega_{\text{r}}$.
From this consideration of parameters, we can expand the term $\cos\left(k_{\delta}z_{j}+\mu_{\delta}t+\phi_{j}\right)$
with small parameters $\eta_{\text{cm}},\eta_{\text{r}}$, but $\eta_{\text{mm}}$
is a big term which needs to be treated exactly. After the expansion,
to leading order in $\eta_{\text{cm}}$ and $\eta_{\text{r}}$, the
Hamiltonian $H$ takes the form 
\begin{equation}
H\approx-\left[\chi_{1}(t)\sigma_{1}^{z}+\chi_{2}(t)\sigma_{2}^{z}\right]\hat{f}_{\text{cm}}-\left[\chi_{1}(t)\sigma_{1}^{z}-\chi_{2}(t)\sigma_{2}^{z}\right]\hat{f}_{\text{r}},\label{9}
\end{equation}
where we have defined 
\begin{eqnarray}
\hat{f}_{\mu} & \equiv & \eta_{\mu}\left(v_{_{\mu}}^{\ast}(t)\hat{a}_{\mu}+v_{\mu}(t)\hat{a}_{\mu}^{\dagger}\right),\\
\chi_{j}(t) & \equiv & \hbar\Omega_{j}\sin\left[\mu_{\delta}t+\phi_{j}-(-1)^{j}\eta_{\text{mm}}\left(t\right)\right],
\end{eqnarray}
where the subscript $\mu=$ cm, r and $j=1,\,2$. In Eq. (9), we have
dropped the term $\cos\left(\mu_{\delta}t+\phi_{j}\pm\eta_{\text{mm}}\right)$
which induces single-bit phase shift but is irrelevant for the CPF\ gate.
The evolution operator at the gate time $\tau$ generated by the Hamiltonian
$H$ can be expressed as 
\begin{equation}
U(\tau)=D_{\text{cm}}(\alpha_{\text{cm}})D_{\text{r}}(\alpha_{\text{r}})\exp\left[i(\gamma_{\text{r}}-\gamma_{\text{\text{cm}}})\sigma_{1}^{z}\sigma_{2}^{z}\right],\label{11}
\end{equation}
where the displacement operator $D_{\mu}(\alpha_{\mu})\equiv\exp\left(\alpha_{\mu}\hat{a}_{\mu}^{\dagger}-\alpha_{\mu}^{\ast}\hat{a}_{\mu}\right)$
($\mu=$ cm, r). Let $j_{\mu}=1$ for $\mu=$ cm and $j_{\mu}=-1$
for $\mu=$ r. The displacement $\alpha_{\mu}$ and the accumulated
phase $\gamma_{\mu}$ have the following expression 
\begin{align}
\alpha_{\mu} & =i\eta_{\mu}\int_{0}^{\tau}\left(\chi_{1}(t)\sigma_{1}^{z}+j_{\mu}\chi_{2}(t)\sigma_{2}^{z}\right)u_{\mu}(t)\, dt\label{12}\\
\gamma_{\mu} & =i\left(\eta_{\mu}\right)^{2}\int_{0}^{\tau}dt_{1}\int_{0}^{t_{1}}dt_{2}\mathbb{\mathcal{S}}[\chi_{1}\chi_{2}]\text{Im}\left[u_{\mu}(t_{1})u_{\mu}^{\ast}(t_{2})\right]
\end{align}
where $\mathbb{\mathcal{S}}[\chi_{1}\chi_{2}]\equiv\chi_{1}(t_{1})\chi_{2}(t_{2})+\chi_{1}(t_{2})\chi_{2}(t_{1})$.

\begin{figure}

\begin{centering}
\includegraphics{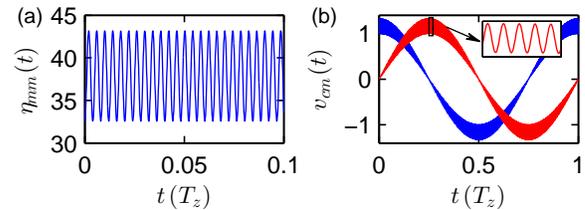}\caption{(Color online) (a) The time dependent parameter $\eta_{mm}(t)$ and
(b) the function $v_{\text{cm}}(t)$. The real/imaginary part (blue/red
curves) of $v_{\text{cm}}(t)$ has even/odd parity as a function of
time and looks similar to $\cos/\sin$ function. Unit of time is the
trap frequency $T_{z}=2\pi/\omega_{cm}$. The parameters used are:
ion mass $m=9u$ ($u$ is the atomic mass unit) corresponding to $Be^{+}$;
r.f. trap frequency $\Omega_{T}=2\pi\times240$MHz; the characteristic
electrode size $d_{0}=200$ $\upmu\text{m}$; AC/DC voltages $V_{0},\, U_{0}$
are $300$V and $21$V respectively. The resulting secular trap frequencies
are $\omega_{cm}=2\pi\times0.965\text{MHz }$, $\omega_{r}=2\pi\times3.62\text{MHz}$
along the $z$-axis, and $\omega_{x}=\omega_{y}=2\pi\times20.8\text{MHz}$
along $x$- and $y$-axis.\label{fig:trajectory}}

\par\end{centering}

\end{figure}

To realize the CPF gate, we require $\alpha_{\mu}=0$ and $\gamma_{\text{r}}-\gamma_{\text{\text{cm}}}=\pi/4$.
The integrals $\alpha_{\mu}$ can be evaluated semi-analytically \cite{supplementary}
or purely numerically. We normally take $\Omega_{1}=\Omega_{2}\equiv\Omega$.
Note that even in this case $\chi_{1}(t_{1})\neq\chi_{2}(t_{2})$
with the micromotion term $\eta_{\text{mm}}\left(t\right)$. This
is different from the case of a static trap. From Eq. (12), we see
that $\alpha_{\mu}=0$ for a fixed $\mu$ gives two complex and thus
four real constraints. With excitation of $N$ motional modes, the
total number of (real) constraints to realize the CPF\ gate is therefore
$4N+1$ (the condition $\gamma_{\text{r}}-\gamma_{\text{\text{cm}}}=\pi/4$
gives one constraint). To satisfy these constraints, we divide the
Rabi frequency $\Omega\left(t\right)$ $\left(0\leq t\leq\tau\right)$
into $m$ equal-time segments, and take a constant $\Omega_{\beta}$
$\left(\beta=1,2,\cdots,m\right)$ for the $\beta$th segment \cite{zhu_arbitrary-speed_2006,zhu_trapped_2006}.
This kind of modulation can be conveniently done through an acoustic
optical modulator in experiments \cite{choi_optimal_2014}. The Rabi
frequencies are our control parameters. For the two ion case, under
fixed detuning $\mu_{\delta}$ and gate time $\tau$, in general we
can find a solution for the CPF\ gate with $m=9$ segments. For some
specific detuning $\mu_{\delta}$ very close to a secular mode frequency,
off-resonant excitations become negligible and a solution is possible
under one segment of pulse by tuning of the gate time $\tau$, which
corresponds to the case of the S\o{}rensen-M\o{}lmer gate \cite{3}
generalized to include the micromotion correction.

\begin{figure}
\begin{centering}
\includegraphics{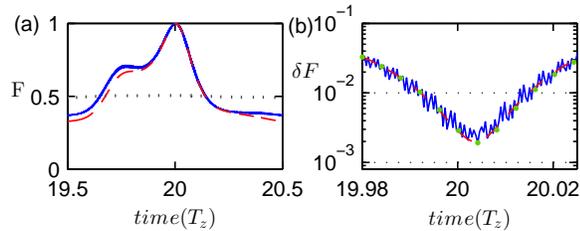}
\par\end{centering}

\caption{(Color online) (a) The fidelity of a two-ion conditional phase flip
gate as a function of gate time, with the unit of time $T_{z}=2\pi/\omega_{cm}$.
The detuning was chosen to be $\mu=0.95\omega_{\text{cm}}$. Blue
solid line indicates the optimal results with micromotion taken into
account; red dashed line is the result for a genuine static harmonic
trap without micromotion; gray dotted line is obtained by applying
the optimal solution for a static harmonic trap to the case with micromotion,
which results in poor performance. (b) The infidelity ($\delta F\equiv1-F$)
near the optimal evolution time, essentially a zoom-in of panel (a)
near $t=20T_{z}$. Green dots in (b) show the time points that are
an integral multiple of the micromotion period. Other parameters used
are: Doppler temperature for both motional degrees of freedom with
$k_{B}T_{D}\approx10\hbar\omega_{\text{cm}}$; effective laser wave
vector $\Delta k=8\upmu\text{m}^{-1}$ so $\eta_{\text{cm}}\approx0.12$
and $\eta_{\text{r}}\approx0.09$.\label{fig:fidelity_mu0.9}}

\end{figure}

To characterize the quality of the gate, we use the fidelity $F\equiv tr_{\mu}\left[\rho_{\mu}\left\vert \left\langle \Psi_{0}\right\vert U_{\text{CPF}}^{\dagger}U(\tau)\left\vert \Psi_{0}\right\rangle \right\vert ^{2}\right]$,
defined as the overlap of the evolution operator $U(\tau)$ with the
perfect one $U_{\text{CPF}}\equiv e^{i\pi\sigma_{1}^{z}\sigma_{2}^{z}/4}$
under the initial state $\left\vert \Psi_{0}\right\rangle $ for the
ion spins and the thermal state $\rho_{\mu}$ for the phonon modes.
In our calculation, without loss of generality, we take $\left\vert \Psi_{0}\right\rangle =\left(\left\vert 0\right\rangle +\left\vert 1\right\rangle \right)\otimes\left(\left\vert 0\right\rangle +\left\vert 1\right\rangle \right)/2$
and assume the Doppler temperature $T_{D}$ for all the phonon modes.
For any given detuning $\mu_{\delta}$ and gate time $\tau$, we optimize
the control parameters $\Omega_{\beta}$ $\left(\beta=1,2,\cdots,m\right)$
to get the maximum fidelity $F$. In Fig. (\ref{fig:fidelity_mu0.9}),
we show the gate fidelity as a function of gate time for $\mu_{\delta}=0.95\omega_{\text{cm}}$
(close to a secular frequency) by applying a single segment laser
pulse of a constant Rabi frequency $\Omega$. In the figure, the dashed
line corresponds to the result in a static harmonic trap with the
same secular frequencies but no micromotion. If we take into account
the micromotion contribution but do not change the gate design, the
result is described by the dash-dot line, with a low fidelity about
only $50\%$. When we optimize the gate design (optimize $\Omega_{\beta}$)
including the micromotion correction, the gate fidelity is represented
by the solid line, which approaches the optimal fidelity achievable
in a static trap. The gate infidelity $\delta F\equiv1-F$ approaches
$2\times10^{-3}$ at the optimal gate time $\tau=20.005T_{z}$, where
$T_{z}\equiv2\pi/\omega_{\text{cm}}$.

By applying $9$ segments of laser pulses with optimized $\Omega_{\beta}$
$\left(\beta=1,2,\cdots,9\right)$, the gate fidelity $F$ can attain
the unity at arbitrary detuning $\mu_{\delta}$ for the two ion case.
As an example, In Fig. 3(a), we show the optimized solution of $\Omega_{\beta}$
(blue lines) at an arbitrarily chosen detuning $\mu_{\delta}=1.4\omega_{\text{cm}}$.
For comparison, the red lines represent the solution of $\Omega_{\beta}$
in a static harmonic trap with otherwise the same parameters. The
maximum magnitude of $\left\vert \Omega_{\beta}\right\vert $ significantly
increases in the presence of micromotion. This is understandable as
fast oscillations of the micromotion tend to lower the effective Rabi
frequencies. In Fig. 3(b), we show the maximum magnitude of $\left\vert \Omega_{\beta}\right\vert $
as a function of the gate time $\tau$. Compared with the solution
in a static harmonic trap, the maximum $\left\vert \Omega_{\beta}\right\vert $
in general needs to increase by about an order of magnitude under
micromotion. 

\begin{figure}

\begin{centering}
\includegraphics{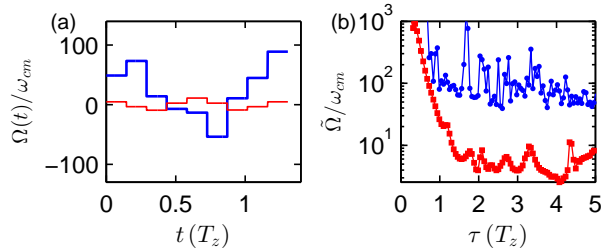}\caption{(Color online) (a) The waveform of the optimal segmented pulse calculated
for the gate with duration $\tau=1.31T_{z}$. The thick blue (thin
red) line corresponds to the case with (without) micromotion. (b)
The maximal Rabi frequency $\tilde{\Omega}\equiv\max_{t}\left|\Omega(t)\right|$
as a function of the gate time $\tau$. The upper blue (lower red)
curve corresponds to the case with (without) micromotion. \label{fig:segmented}}

\par\end{centering}

\end{figure}

In conclusion, we demonstrate that arbitrarily high fidelity quantum
gates can be achieved under large micromotion. The demonstration in
this paper uses the example of two ions in a quadrupole trap, which
has the micromotion magnitude significantly beyond the Lamb-Dicke
limit. Apparently, the idea here is applicable to the many ion case.
For a system of $N$ ions in any dimension, as long as the ions crystallize,
each ion has an average equilibrium position. We can then expand the
Coulomb potential around these equilibrium positions. Under the r.f.
Paul trap and the Coulomb interaction, the motion of the ions can
then be described by a set of coupled time-dependent Mathieu equations.
Using the technique in this paper, we can solve the motional dynamics
and optimize the gate design that explicitly takes into account all
the micromotion contributions. The gate design technique under micromotion
proposed in this paper solves a major obstacle for high fidelity quantum
computation in real r.f. traps beyond the 1D limitation and opens
a new way for scalable quantum computation based on large 2D or 3D
trap-ion crystals in Paul traps.

\textit{Acknowledgments. }This work was supported by the NBRPC (973
Program) 2011CBA00300 (2011CBA00302), the IARPA MUSIQC program, the
ARO and the AFOSR MURI programs, and the DARPA OLE program.

\begin{widetext}

\section*{Appendix: Supplementary Material}

In this appendix, we show in detail how to solve the driven Mathieu
equation and give an approximate treatment of the motional integrals.

\subsection*{SOLUTION OF DRIVEN MATHIEU EQUATION}

We show in detail how to solve the Mathieu equation with a constant
drive term. 
\[
\frac{d^{2}u}{d\xi^{2}}+\left(a-2q\cos\left(2\xi\right)\right)u=f_{0}
\]
 Let us assume that $u(\xi)=f_{0}\sum_{n=0}^{\infty}c_{n}\cos(2n\xi)$
and insert it into the equation. After re-organization, we get
\[
ac_{0}-qc_{1}+\sum_{n=1}^{\infty}\left[(a-4n^{2})c_{n}-q(c_{n-1}+c_{n+1})-qc_{0}\delta_{n,1}\right]\cos(2nt)=1.
\]
 Defining $D_{n}\equiv(a-4n^{2})/q$, we have the following set of
linear equations
\begin{eqnarray*}
ac_{0}-qc_{1} & = & 1\\
c_{n}-\frac{1}{D_{n}}(c_{n-1}+c_{n+1}+c_{0}\delta_{n,1}) & = & 0.
\end{eqnarray*}
 In matrix form, 
\begin{equation}
\left(\begin{array}{cccccc}
a & -q & 0 & \cdots &  & 0\\
-\frac{2}{D_{1}} & 1 & -\frac{1}{D_{1}}\\
0 & -\frac{1}{D_{2}} & 1 & -\frac{1}{D_{2}}\\
\vdots &  & -\frac{1}{D_{3}} & 1 & -\frac{1}{D_{3}}\\
 &  &  & \ddots & \ddots\\
0
\end{array}\right)\cdot\left(\begin{array}{c}
c_{0}\\
c_{1}\\
c_{2}\\
\vdots\\
\\
\\
\end{array}\right)=\left(\begin{array}{c}
1\\
0\\
0\\
\vdots\\
\\
\\
\end{array}\right).\label{eq:matrix_eq}
\end{equation}
The factor $1/D_{n}$ decreases very fast as $n$ increases and we
can truncate the expansion of $u(\xi)$ at a small $n$. Numerically
we observe that typically keeping up to $c_{2}$ already gives enough
accuracy. We can thus get a very accurate analytical expression 
\begin{eqnarray*}
c_{0} & \approx & \frac{64+a(a-20)-q^{2}}{(32-3a)q^{2}+a(a-4)(a-16)},\\
c_{1} & \approx & \frac{2(a-16)q}{(32-3a)q^{2}+a(a-4)(a-16)},\\
c_{2} & \approx & \frac{2q^{2}}{(32-3a)q^{2}+a(a-4)(a-16)}.
\end{eqnarray*}
For the example in the main text, $a_{r}=-0.0388$ and $q_{r}=0.283$,
we have $c_{0}=1132.8$ and $u_{r}(\xi)=f_{0}c_{0}\left[1-0.14\cos(2\xi)+0.0025\cos(4\xi)+\cdots\right].$ 

The micromotion corrected equilibrium position is $f_{0}c_{0}$ and
should be identified with $u_{0}$ around which we expand the Coulomb
potential in the first place. Thus we should determine them self-consistently.
Taking the relative motion in the manuscript as an example, since
both $a_{\text{r}}\equiv\frac{-16eU_{0}}{md_{0}^{2}\Omega_{T}^{2}}+\frac{4e^{2}}{\pi\epsilon_{0}mu_{o}^{3}\Omega_{T}^{2}}$
and $f_{0}\equiv\frac{6e^{2}}{\pi\epsilon_{0}mu_{0}^{2}\Omega_{T}^{2}}$
are functions of $u_{0}$, then the self-consistent equation 
\begin{eqnarray*}
u_{0}=f_{0}c_{0} & \approx & f_{0}\frac{64+a_{\text{r}}(a_{\text{r}}-20)-q_{\text{r}}^{2}}{(32-3a_{\text{r}})q_{\text{r}}^{2}+a_{\text{r}}(a_{\text{r}}-4)(a_{\text{r}}-16)}
\end{eqnarray*}
 gives the  correct $u_{0}$. With the iterative method it typically
takes only a few iterations to converge to the correct value when
starting from a proper initial value of $u_{0}$.

\subsection*{TWO-STAGE TIME INTEGRAL}

Here we offer an approximate treatment of motional integrals. We notice
that the secular frequency $\omega$ and the micromotion frequency
$\Omega$ are well separated, i.e. $\omega\ll\Omega$. This means
quantities with characteristic frequency $\omega$ or below stay constant
within one period of micromotion. So we can perform the time integral
in two steps: we first integrate over one period of the micromotion,
obtaining a slowly varying integrand, which we then integrate again.
By doing this we will show that the dominant effect of micromotion
is to modulate the effective Rabi frequency. Notice that the integrals
$\int_{0}^{\tau}\chi(t)u(t)\, dt$ can be reduced to the form below
(ignoring micromotion frequencies $n\Omega\pm\omega$ with $n\ge2$)
\begin{eqnarray*}
I & = & \int_{0}^{\tau}\sin\left(a_{0}(t)+a_{1}(t)\cos(\Omega t_{1}+\phi(t))\right)\left(b_{0}(t)+b_{1}(t)\cos\left(\Omega t+\varphi(t)\right)\right)
\end{eqnarray*}
 where $a_{0}(t)$, $a_{1}(t)$, $b_{0}(t)$, $b_{1}(t)$, $\phi(t)$
and $\varphi(t)$ are all real slowly varying functions within one
period of micromotion $\frac{2\pi}{\Omega}$. The above integral can
be further broken into two parts, $I_{1}$ and $I_{2}$, where 

\begin{eqnarray*}
I_{1} & \approx & \int_{0}^{\tau}dt\frac{\Omega}{2\pi}\int_{t}^{t+2\pi/\Omega}dt_{1}\sin\left(a_{0}(t)+a_{1}(t)\cos(\Omega t_{1}+\phi)\right)b_{0}(t)\\
 & = & \int_{0}^{\tau}dt\,\frac{1}{2\pi}\int_{-\pi}^{\pi}dt'\,\sin\left(a_{0}(t)+a_{1}(t)\cos(t')\right)b_{0}(t)\\
 & = & \text{Im}\left[\int_{0}^{\tau}dt\,\exp\left(i\, a_{0}(t)\right)\frac{1}{2\pi}\int_{-\pi}^{\pi}dt'\,\exp\left(i\, a_{1}\cos(t')\right)b_{0}(t)\right]\\
 &  & \text{Im}\left[\int_{0}^{\tau}dt\,\exp\left(i\, a_{0}(t)\right)J_{0}(a_{1})b_{0}(t)\right]\\
 & = & \int_{0}^{\tau}dt\,\sin\left(a_{0}(t)\right)b_{0}(t)J_{0}(a_{1}(t))
\end{eqnarray*}
and 
\begin{eqnarray*}
I_{2} & = & \int_{0}^{\tau}dt\sin\left(a_{0}(t)+a_{1}(t)\cos(\Omega t+\phi)\right)b_{1}(t)\cos\left(\Omega t+\varphi(t)\right)\\
 & \approx & \int_{0}^{\tau}dt\frac{\Omega}{2\pi}\int_{t}^{t+2\pi/\Omega}dt_{1}\sin\left(a_{0}(t)+a_{1}(t)\cos(\Omega t_{1}+\phi)\right)b_{1}(t)\cos\left(\Omega t_{1}+\varphi\right)\\
 & = & \int_{0}^{\tau}dt\,\cos(a_{0}(t))\cos(\varphi-\phi)J_{1}(a_{1}(t))
\end{eqnarray*}
where $J_{0}$ and $J_{1}$ denote the Bessel functions. In both cases,
the micromotion gives rise to slowly varying modulation factors, $J_{0}(a_{1}(t))$
and $\cos(\varphi-\phi)J_{1}(a_{1}(t))$. Moreover in $I_{2}$ the
phase of the original integrand is also shifted, $\sin(a_{0}(t))\rightarrow\cos(a_{0}(t))$.
For the actual experimental system, the term $I_{2}$ contributes
much less than the $I_{1}$ to the target integral $I$, due to the
much smaller coefficient of the micromotion component than that of
the secular component in $v(t)$. So in leading order, micromotion
reduces the laser Rabi frequency seen by the ion by a factor on the
order of $J_{0}(a_{1}(t))$. 

\end{widetext}


\begin{thebibliography}{10}
\bibitem{1}For a review, see D. Leibfried, R. Blatt, C. Monroe, D.
Wineland, Rev. Mod. Phys. 75, 281-324 (2003); R. Blatt and D. Wineland,
Nature 453, 1008-1015 (2008); C. Monroe, J. Kim, Science 339, 1164-1169
(2013).

\bibitem{2}J. I. Cirac and P. Zoller, Phys. Rev. Lett. 74, 4091-4094
(1995). 

\bibitem{3}A. S\o{}rensen and K. M\o{}lmer, Phys. Rev. Lett. 82,
1971 (1999); G. J. Milburn, S. Schneider, and D. F. V. James, Fortschr.
Physik 48, 801-810 (2000); A. S\o{}rensen, K. M\o{}lmer, Phys. Rev.
A 62, 022311 (2000).

\bibitem{4}C. A. Sackett, et. al., Nature 404, 256 (2000); D. Liebfried
et al., Nature 422, 412-415 (2003); F. Schmidt-Kaler, et. al., Nature
422, 408 (2003); H. H\"{a}ffner, et.al, Nature, 438, 643 (2005);
J. Benhelm, G. Kirchmair, C. F. Roos, and R. Blatt, Nature Physics
4, 463 (2008); K. Kim, et. al., Nature 465, 590 (2010); R. Islam,
et al., Nature Comm. 2, 377 (2011); B. P. Lanyon, et. al., Science
334, 57 (2011); B. P. Lanyon, et. al., Phys. Rev. Lett. 111, 210501
(2013); R. Islam, et. al., Science 340, 583 (2013). 

\bibitem{5}M. G. Raizen, J. M. Gilligan, J. C. Bergquist, W. M. Itano,
and D. J. Wineland, Phys. Rev. A 45, 6493 (1992); J. P. Schiffer,
Phys. Rev. Lett. 70, 818 (1993); G.-D. Lin, et al., Europhys. Lett.
86, 60004 (2009).

\bibitem{6}D. Gottesman, J. Mod. Opt. 47, 333-345 (2000); T. Szkopek
et al., IEEE Trans. Nano., Vol. 5, No. 1, pp 42-49, 2006. 

\bibitem{raussendorf_fault-tolerant_2007}R. Raussendorf and J. Harrington,
Phys. Rev. Lett. 98, 190504 (2007). 

\bibitem{2D_crystal}A. Mortensen, E. Nielsen, T. Matthey, and M.
Drewsen, Phys. Rev. Lett. 96, 103001 (2006); K. Okada, T. Takayanagi,
M.Wada, S. Ohtani, and H. A. Schuessler, Phys. Rev. A 80, 043405 (2009);
B. Szymanski, et. al., App. Phys. Lett. 100, 171110 (2012); M. Drewsen,
T. Matthey, A. Mortensen, and J. P. Hansen, arXiv:1202.2544 (2012).

\bibitem{zhu_arbitrary-speed_2006}S.-L. Zhu, C. Monroe, and L.-M.
Duan, Europhys. Lett. 73, 485 (2006).

\bibitem{zhu_trapped_2006}S.-L. Zhu, C. Monroe, and L.-M. Duan, Phys.
Rev. Lett. 97, 050505 (2006).

\bibitem{mclachlan_mathieu_1947}N. W. McLachlan, Theory and Application
of Mathieu Functions (Clarendon Press, Oxford, 1947).

\bibitem{ref_oscillator}M. Combescure, Annales de l'institut Henri
Poincare (A) Physique theorique 44, 293 (1986); L. S. Brown, Phys.
Rev. Lett. 66, 527 (1991); R. J. Glauber, in Laser Manipulation of
Atoms and Ions (1922), vol. 118 of Proceedings of the International
School of Physics \textquotedbl{}Enrico Fermi\textquotedbl{} Course;
B. E. King, Ph.D. thesis, University of Colorado at Boulder (1999).

\bibitem{supplementary} See supplementary material in the appendix
for details of the solution of the inhomogeneous Mathieu equation
and an approximate treatment of the motional integrals. 

\bibitem{choi_optimal_2014}T. Choi, S. Debnath, T. A. Manning, C.
Figgatt, Z.- X. Gong, L.-M. Duan, and C. Monroe, arXiv:1401.1575 (2014).\end{thebibliography}
\end{document}